\documentclass[prd,aps,12pt,nofootinbib,superscriptaddress]{revtex4-1}
\usepackage{amsmath, amssymb, amsfonts, amsthm}
\usepackage[usenames,dvipsnames,svgnames,table]{xcolor}
\usepackage{graphicx}
\usepackage{subfigure}

\usepackage{epsfig}

\def\f{\frac}
\def\l{\left}
\def\r{\right}
\def\ie{{\it i.e.~}}
\def\beq{\begin{equation}}
\def\eeq{\end{equation}}
\def\ber{\begin{eqnarray}}
\def\eer{\end{eqnarray}}

\def \lleq {\lower0.9ex\hbox{ $\buildrel < \over \sim$} ~}
\def \ggeq {\lower0.9ex\hbox{ $\buildrel > \over \sim$} ~}

\def\prl{{Phys.\@ Rev.\@ Lett.\ }}
\def\prd{{Phys.\@ Rev.\@ D\ }}

\def\plb {{Phys.\@ Lett.\@ B\ }}

\def\etal{{\it et al.}}

\begin{document}

\title{Unifying Dark Matter and Dark Energy with non-Canonical Scalars}

\author{Swagat S. Mishra and Varun Sahni}
\affiliation{Inter-University Centre for Astronomy and Astrophysics,
Post Bag 4, Ganeshkhind, Pune 411~007, India}

\thispagestyle{empty}

\sloppy

\begin{abstract}
Non-canonical scalar fields with the Lagrangian ${\cal L} = X^\alpha - V(\phi)$,
possess the attractive property that the speed of sound, $c_s^{2} = (2\,\alpha - 1)^{-1}$,
can be exceedingly small for large values of $\alpha$.
This allows a non-canonical field to cluster and behave like warm/cold dark matter on small scales.
We demonstrate that simple potentials including $V = V_0\coth^2{\phi}$ and
a Starobinsky-type potential
can unify dark matter and
dark energy. 
{\em Cascading dark energy}, in which the potential cascades to lower values
in a series of discrete steps, can also work as a unified model.
In all of these models the kinetic term $X^\alpha$ plays the role of dark matter,
while the potential term $V(\phi)$ plays the role of dark energy.

\end{abstract}

\maketitle


\bigskip

\section{Introduction}
\label{sec:intro}

A key feature of our universe is that $96\%$ of its matter content is
weakly interacting and non-baryonic. 
It is widely believed that this so-called dark sector consists of two distinct sub-components,
the first of which, dark matter (DM), consists of a pressureless fluid which clusters,
while the second, dark energy (DE), has large negative pressure and causes the universe to accelerate
at late times.

Although numerous theoretical models have been advanced as to what may constitute dark matter,
none so far has received unambiguous experimental support \cite{dark_matter}.
The same may also be said of dark energy.
The simplest model of DE, the cosmological constant $\Lambda$, fits most observational data sets quite well
\cite{planck_2015}; see however \cite{bao_2014,sss14}. Yet
the fine tuning problem  associated with $\Lambda$ and the cosmic coincidence issue, have motivated the
development of dynamical dark energy (DDE) models in which the DE density and equation of state (EOS)
evolve with time \cite{review_DE,sahni04}.

In view of the largely unknown nature of the dark sector several
 attempts have been made to describe it within
a unified setting \cite{CG,CG1,tirth,CG_pert,sandvik04,scherrer04,tejedor_feinstein,bertacca07,fang07,bertacca08,bertacca11,cervantes11,asen14,sahni_sen,li-scherrer,gurzadyan}. Perhaps the earliest prescription for a unified model of dark matter and dark energy
was made in the context of the Chaplygin gas (CG) \cite{CG,CG1,CG_pert}. CG possesses an equation of state
which is pressureless ($p \simeq 0$) at early times and $\Lambda$-like ($p \simeq -\rho$)
at late times. This led to the hope that CG may be able to describe both dark matter and dark energy
through a unique Lagrangian.
Unfortunately an analysis of density perturbations dashed these early hopes for unification \cite{sandvik04}.
In this paper we build on early attempts at unification and demonstrate that a compelling unified description
of dark matter and dark energy can emerge from scalar fields with non-canonical kinetic terms.

\section{Non-canonical scalar fields}
\label{sec:NC}

The non-canonical scalar field Lagrangian
\cite{Mukhanov-2006,sanil-2008,sanil13}
\beq
{\cal L}(X,\phi) = X\left(\frac{X}{M^{4}}\right)^{\alpha-1} -\; V(\phi),
\label{eqn: Lagrangian}
\eeq
presents a simple generalization of the canonical scalar field Lagrangian
\beq
{\cal L}(X,\phi) = X -\; V(\phi), ~~~~X = \frac{1}{2}{\dot\phi}^2
\label{eqn: Lagrangian0}
\eeq
to which (\ref{eqn: Lagrangian}) reduces when $\alpha=1$.

Two properties of non-canonical scalars make them attractive for the study of 
cosmology:
\begin{enumerate}

\item Their equation of motion
\beq
{\ddot \phi}+ \f{3\, H{\dot \phi}}{2\alpha -1} + \left(\f{V'(\phi)}{\alpha(2\alpha -1)}\right)\left(\f{2\,M^{4}}{{\dot \phi}^{2}}\right)^{\alpha - 1} =\; 0,
\label{eqn: EOM-model}
\eeq
 is of second order, as in the canonical case.
Indeed, (\ref{eqn: EOM-model}) reduces to the standard canonical form
${\ddot \phi}+ 3\, H {\dot \phi} + V'(\phi) = 0$ when $\alpha =1$.

\item The speed of sound \cite{Mukhanov-2006}
\beq
c_s^{2} = \f{1}{2\,\alpha - 1}
\label{eqn: sound speed model}
\eeq
can become quite small for large values of $\alpha$
since $c_s \to 0$ when $\alpha \gg 1$.
\end{enumerate}

This latter property ensures that non-canonical scalars can, in principle,
 play the role of dark matter.
In this paper we shall show that, for a suitable choice of the potential $V(\phi)$,
non-canonical scalars can also unify dark matter with dark energy.

This paper works in the context of a spatially flat  Friedmann-Robertson-Walker (FRW) universe
for which the energy-momentum tensor has the form
\beq
T^{\mu}_{\;\:\;\nu} = \mathrm{diag}\left(\rho_{_{\phi}}, -p_{_{\phi}}, - p_{_{\phi}}, - p_{_{\phi}}\r).
\eeq
In the context of non-canonical scalars,
 the energy density, $\rho_{_{\phi}}$, and pressure, $p_{_{\phi}}$, can be written as
\ber
\rho_{_{\phi}} &=& \l(2\alpha-1\r)X\l(\frac{X}{M^{4}}\r)^{\alpha-1} +\;  V(\phi),\nonumber\\
p_{_{\phi}} &=& X\l(\frac{X}{M^{4}}\r)^{\alpha-1} -\; V(\phi).
\label{eqn:rho}
\eer
It is easy to see that (\ref{eqn:rho}) reduces to the the canonical form
$\rho_{_{\phi}} = X + V$, ~$p_{_{\phi}} = X - V$ when $\alpha = 1$.

The Friedmann equation which is solved in association with \eqref{eqn: EOM-model} is
\beq
H^{2} = \frac{8 \pi G}{3}\l(\rho_{_{\phi}} + \rho_{_r} +\rho_{_b}\r)
\label{eqn:freid}
\eeq
where $\rho_{_r}$ is the radiation term and $\rho_{_b}$ is the contribution from baryons.
Note that we do not assume a separate contribution from dark matter or dark energy which are encoded
in $\rho_{_{\phi}}$.

As shown in \cite{sahni_sen}, for sufficiently flat potentials with $V' \simeq 0$
the third term in (\ref{eqn: EOM-model}) can be neglected, leading to
\beq
{\ddot \phi} \simeq -\f{3\, H{\dot \phi}}{2\alpha -1}~~~ \Rightarrow
~~{\dot\phi} \propto a^{-\frac{3}{2\alpha - 1}}~.
\label{eq:phidot}
\eeq
Substituting (\ref{eq:phidot}) in 
\beq
\rho_{_X} = \l(2\alpha-1\r)X\l(\frac{X}{M^{4}}\r)^{\alpha-1},~~~~X = \frac{1}{2}{\dot\phi}^2
\label{eqn:rhoNC}
\eeq
one finds
\beq
\rho_{_X} \propto a^{-\frac{6\alpha}{2\alpha - 1}}
\label{eq:kinetic}
\eeq
which reduces to $\rho_{_X} \propto a^{-3}$ for $\alpha \gg 1$.
Comparing \eqref{eq:kinetic} with $\rho_{_X} \propto a^{-3(1+w_{_X})}$ one finds
\beq
w_{_X} = \frac{1}{2\alpha-1}~,
\label{eq:state}
\eeq
so that
$w_{_X} \simeq 0$ for $\alpha \gg 1$.
We therefore conclude that for flat potentials and large values of $\alpha$,
the kinetic term, $\rho_{_X}$, plays the role of dark matter while the potential term, $V$,
plays the role of dark energy in (\ref{eqn:rho}).

The above argument is based on the requirement that the third term
is much smaller than the first two terms in equation (\ref{eqn: EOM-model}).
In a recent paper Li and Scherrer \cite{li-scherrer} have made the interesting
observation that the potential need not be flat in order that
(\ref{eq:phidot}) be satisfied.
The analysis of Li and Scherrer suggests that the third term
in (\ref{eqn: EOM-model}) can be neglected under more general conditions than
anticipated in \cite{sahni_sen}
provided the potential is `sufficiently rapidly decaying' \cite{li-scherrer}.
Indeed it is easy to show that the requirement of the third term in (\ref{eqn: EOM-model}) being much
smaller than the second translates into the inequality
\beq
V' \ll \left (\frac{2\alpha - 1}{2\alpha}\right )\frac{3H\rho_{_X}}{\dot\phi}~,
\label{eq:inequality1}
\eeq
which reduces to
\beq
V' \ll \frac{3H\rho_{_X}}{\dot\phi} ~~\Rightarrow~~ {\dot V} \ll 3H\rho_{_X},
\label{eq:inequality2}
\eeq
when $\alpha \gg 1$.
Equation \eqref{eq:inequality2} can be recast as
\beq
\left\vert \frac{dV}{dz}\right\vert \ll \frac{3\rho_{_X}}{1+z}~.
\label{eq:inequality3}
\eeq
Equations \eqref{eq:inequality2} \& \eqref{eq:inequality3} inform us that the variation in $V$
can be quite large at early times when both $H$ and $\rho_{_X}$ are large. 
By contrast the same equations suggest that the potential should be quite flat at late times
when $H$ and $\rho_{_X}$ are small.
This latter property ensures that $V(\phi)$ can play the role of DE at late times
and drive cosmic acceleration.
The inequalities in \eqref{eq:inequality2}, \eqref{eq:inequality3} can be satisfied
by a number of potentials some of which are discussed below.

\begin{figure}[htb]
\centering
\includegraphics[width=0.85\textwidth]{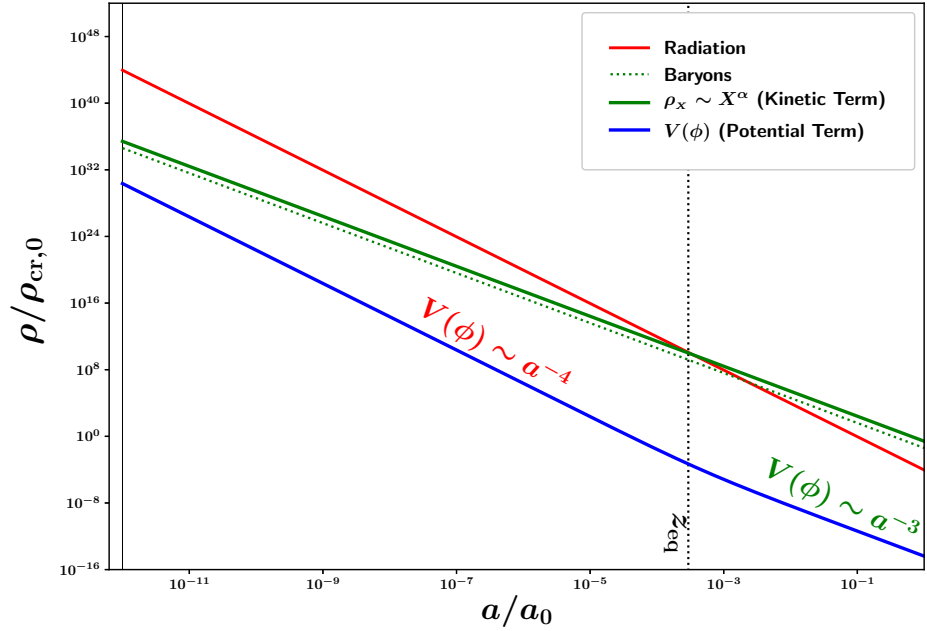}
\caption{\small This figure illustrates  the evolution of the density in radiation (red curve), baryons 
(dotted green curve) and a non-canonical scalar field with the potential $V \propto \phi^{-2}$. 
The solid green curve shows the evolution of the kinetic term $\rho_{_X} \propto X^\alpha$ which behaves like dark matter,
$\rho_{_X} \propto a^{-3}$. The blue curve shows the  evolution of the potential $V(\phi)$. Note that $V$ 
scales like the background fluid 
so that $V \propto a^{-4}$ during the radiative regime and $V \propto a^{-3}$ during matter domination.
($\alpha = 10^5$ has been assumed.)}
\label{fig:rho_ipl}
\end{figure}

We first consider 
the inverse-power-law (IPL) family of potentials \cite{ratra88a}
\beq
V(\phi) = \frac{V_0}{(\phi/m_p)^p},~~p>0.
\label{eq:IPL}
\eeq
Assuming that the background density falls off as
\beq
\rho_{_{\rm B}} \propto a^{-m}
\label{eq:background}
\eeq
(where $m=4, 3$ for radiative and matter dominated epochs respectively),
one can show that (\ref{eq:phidot})
is a late-time attractor provided the inequality
\beq
p \geq \frac{12\alpha}{(2\alpha - 1)m - 6}
\label{eq:inequality}
\eeq
is satisfied \cite{li-scherrer}.
For the large values $\alpha \gg 1$ which interest us,
(\ref{eq:inequality}) reduces to
\beq
p \geq \frac{6}{m}
\eeq
for which the late-time attractor is \cite{li-scherrer}
\beq
V \propto \l(\rho_{_{\rm B}}\r)^\frac{p}{2} ~~~{\rm and} ~~~ \rho_{_X} \propto a^{-3}~.
\label{eq:PE_att}
\eeq
From \eqref{eq:PE_att} one
 finds that for $p > 2$ the potential falls off {\em faster} than the background density $\rho_{_{\rm B}}$.
The case $p=2$ is special since $V \propto \rho_{_{\rm B}}$, \ie
 the potential scales exactly like the background density.
This is illustrated in figure \ref{fig:rho_ipl}.
In this case one finds (for $\alpha \gg 1$)
\beq
\frac{V(\phi)}{\rho_{_{\rm B}}} \simeq \frac{V_0}{\rho_{_{0{\rm B}}}} \left (\frac{H_0 m_p}{M^2}\right )^2
\label{eq:ratio}
\eeq
where $M$ is a free parameter in the non-canonical Lagrangian
\eqref{eqn: Lagrangian}.

The above analysis suggests that the IPL potential $V \propto \phi^{-p}, ~ p\geq 2$ cannot give rise to
cosmic acceleration at late times. Thus while being able to account for dark matter (through $\rho_{_X}$) this
model is unable to provide a unified description of dark matter and dark energy.

However, as we show in the rest of this paper, a unified prescription for dark matter and dark energy
is easily provided by any one of the following potentials\footnote{Perhaps the simplest potential for unification is
$V(\phi) = V_0$ which results in a $\Lambda$CDM cosmology \cite{sahni_sen}.}: (i) $V(\phi) = V_0 \coth^2{\phi}$, 
(ii) the Starobinsky-type potential $V(\phi) = V_0 \left ( 1 - e^{-{\phi}}\right )^{2}$, (iii) 
the step-like potential
$V(\phi) = A + B\tanh{\beta \phi}$.
It is interesting that all of these potentials 
belong to the $\alpha$-attractor family \cite{linde1,linde2}
in the canonical case.

\section{Unified models of Dark Matter and Dark Energy}
\label{sec:unified}

\subsection{Dark Matter and Dark Energy from $V = V_0\coth^2{\phi}$}
\label{sec:coth}

The potential \cite{bms17}
\ber
V(\phi) &=& V_0\, \coth^p{\left(\frac{\phi}{m_p}\right)},~~~p>0 \nonumber \\
\nonumber \\
&\equiv& V_0\, \l (\frac{1 + e^{-2\phi/m_p}}{1 - e^{-2\phi/m_p}}\r )^p
\label{eq:coth}
\eer
can provide a compelling description of dark matter and dark energy on account of
its two asymptotes:
\ber
V(\phi) &\simeq& \frac{V_0}{(\phi/m_p)^p}~, ~~~~~~{\rm for} ~~\lambda\phi \ll m_p\label{eq:IPL1}\\
V(\phi) &\simeq& V_0 ~, ~~~~~~~~~~~~~~~{\rm for} ~~\lambda\phi \gg m_p~.
\label{eq:coth_late}
\eer
As discussed in the previous section,
the IPL asymptote (\ref{eq:IPL1}) 
ensures that the kinetic term behaves like dark matter
$\rho_{_X} \propto a^{-3}$, while $V(\phi)$ scales like the background density (for $p=2$) or faster
(for $p > 2$); see eqn. (\ref{eq:PE_att}).

\begin{figure}[htb]
\centering
\begin{center}
\vspace{0.0cm}
$\begin{array}{@{\hspace{-0.5in}}c@{\hspace{-0.2in}}c}
\multicolumn{1}{l}{\mbox{}} &
\multicolumn{1}{l}{\mbox{}} \\ [-0.10in]
\epsfxsize=3.8in
\epsffile{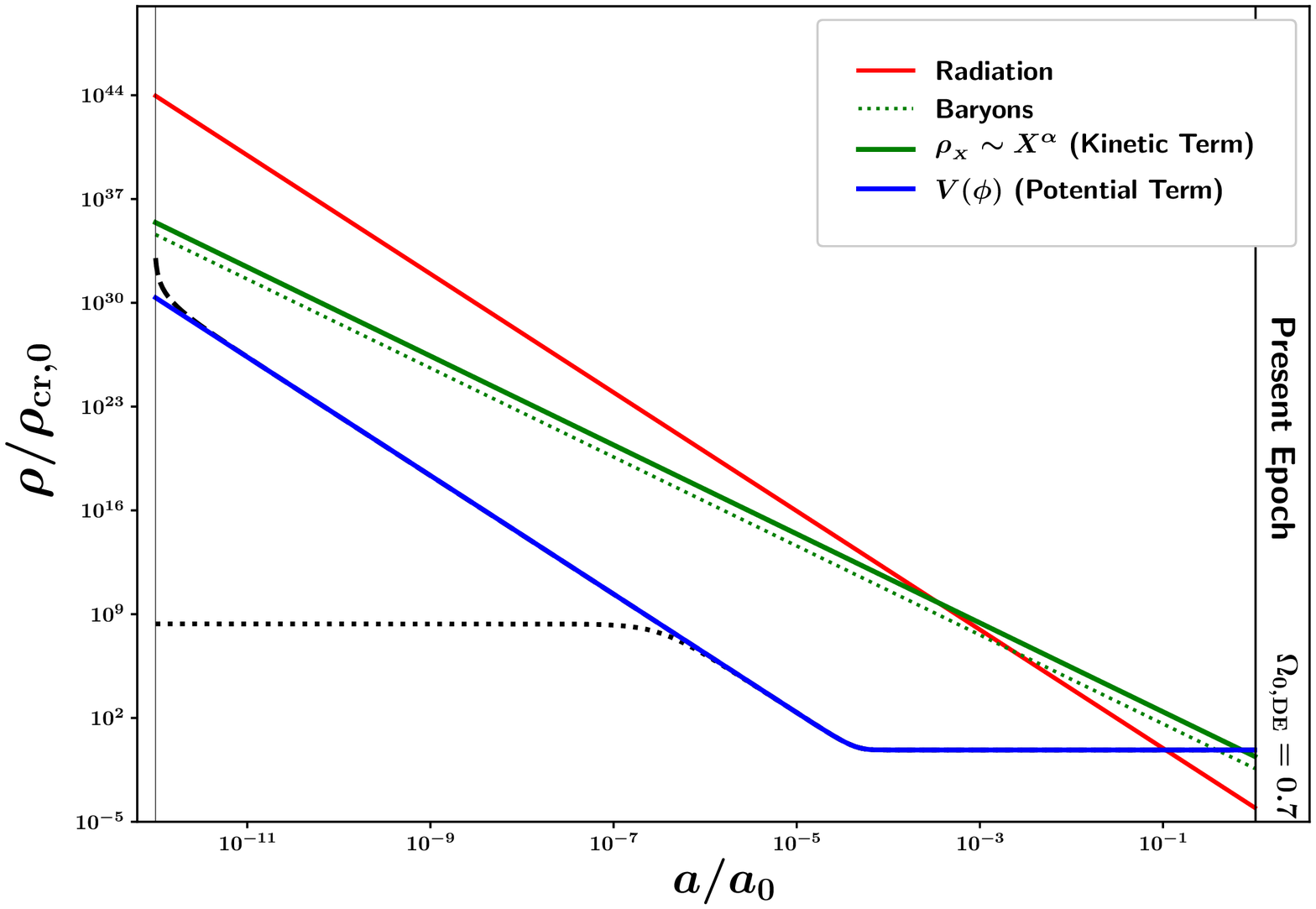} &
\epsfxsize=3.8in
\epsffile{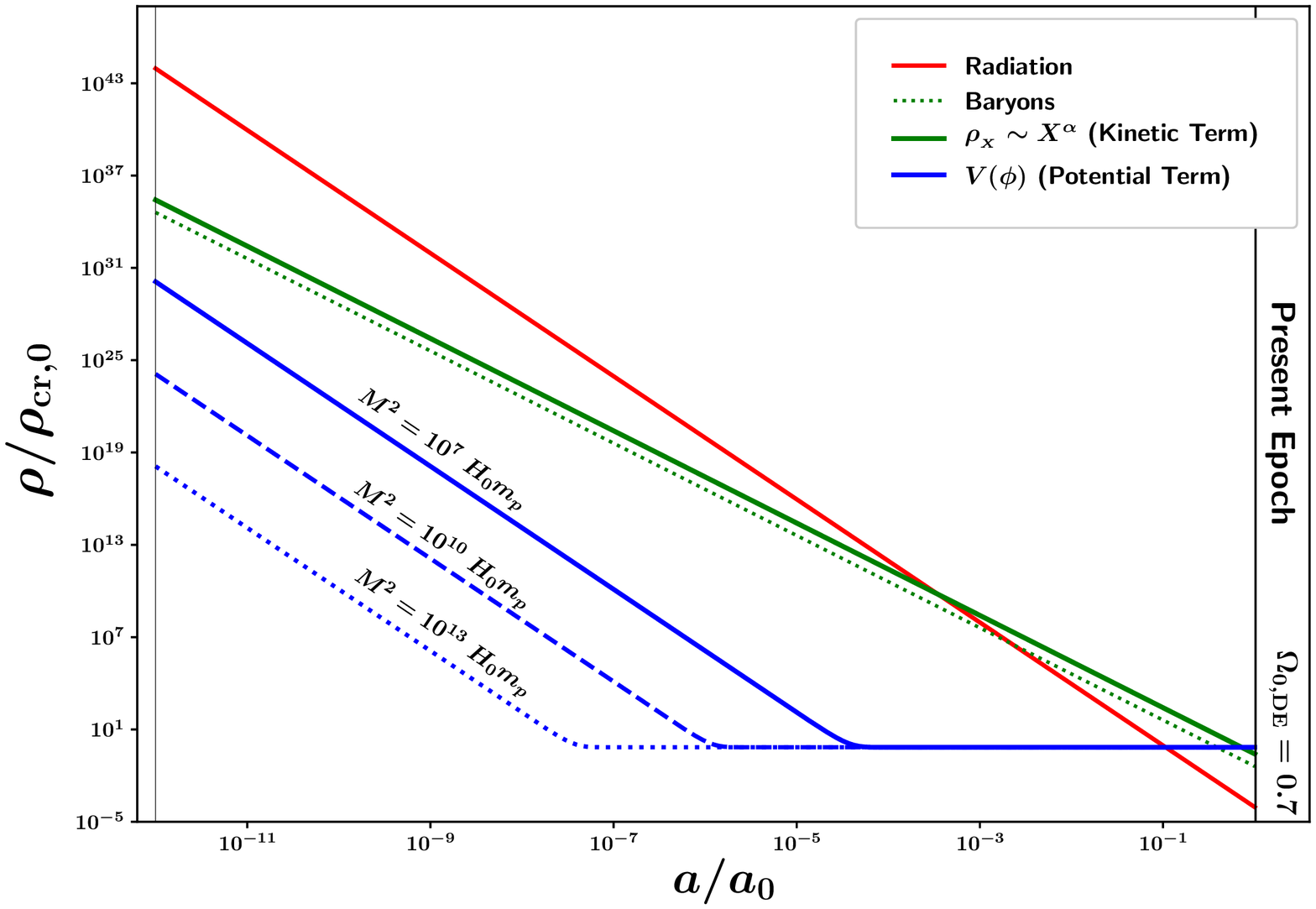} \\
\end{array}$
\end{center}
\vspace{0.0cm}
\caption{\small {\bf Left panel}: The evolution of the density 
(in units of $\rho_{\rm cr,0} = 3 m_p^2 H_0 ^2 $) is shown for radiation (red), baryons (dotted green), kinetic term
$\rho_{_X} \propto X^{\alpha}$ (solid green) and the potential $V(\phi)$ (solid blue) for 
$V(\phi) \propto \coth^{2}{\phi}$. 
Note that dark matter is sourced by the kinetic term which {\em always} drops off as
 $\rho_{_X} \propto a^{-3}$, regardless of the shape of the potential.
DE is sourced by $V(\phi)$
which initially scales like the background fluid, $V \propto \rho_{_B}$, before flattening
to a constant value $V_0$. 
The dotted and dashed black curves correspond to different initial values of $V(\phi)$ 
and indicate that a large range of initial conditions converge onto the scaling attractor (solid blue), yielding $\Omega_{{\rm 0,DE}}=0.7$ at the present epoch.  
{\bf Right panel}: Same as the left panel except that results for three different values
of the non-canonical parameter
$M$ are shown.
Note that dark matter (solid green), sourced by the kinetic term, {\em always} drops off as
 $\rho_{_X} \propto a^{-3}$, regardless of the value of $M$.
By contrast the value of $V(\phi)$ depends quite sensitively on $M$, see \eqref{eq:ratio}.
In all cases $V(\phi)$
initially scales like the background fluid, $V \propto \rho_{_B}$, before flattening
to the constant value $V(\phi) \simeq V_0$. 
Late time expansion in this model resembles $\Lambda$CDM.
}
\label{fig:rho_coth}
\end{figure}

\begin{figure}[htb]
\centering
\includegraphics[width=0.7\textwidth]{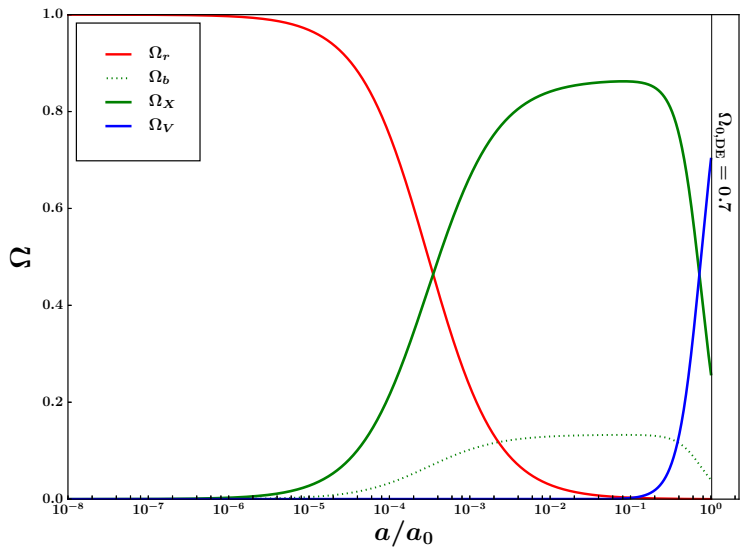}
\caption{\small The evolution of density parameters $(\Omega_i = \rho_i/\rho_{\rm cr})$
is shown for radiation (red), baryons (dotted green), kinetic term
$\rho_{_X} \propto X^{\alpha}$ (solid green) and the potential $V(\phi)$ (solid blue) for 
$V(\phi) \propto \coth^{2}{{\phi}}$. This model gives
rise to a matter dominated epoch (sourced by the kinetic term $\rho_{_X}$) when $z > 1$ and a DE dominated epoch, 
sourced by the potential $V(\phi)$, when $z < 1$.
}
\label{fig:Omega_coth}
\end{figure}

The late-time asymptote (\ref{eq:coth_late}) demonstrates that 
 the potential flattens to a constant
value at late times.  This feature allows $V(\phi)$ to play the role of dark energy.
Indeed, a detailed numerical analysis of the $\coth$ potential, 
summarized in figures \ref{fig:rho_coth} and \ref{fig:Omega_coth}, demonstrates that
(\ref{eq:coth}) with $p = 2$ can 
provide a successful unified description of dark matter and dark energy in the non-canonical setting.
(This is also true for $p > 2$. However for
 shallower potentials with $p < 2$ the third term in (\ref{eqn: EOM-model})
 cannot be neglected. This implies that for such potentials 
$\rho_{_X}$ does not scale as $a^{-3}$ and therefore cannot play the role of dark matter \cite{li-scherrer}.)

\begin{figure}[htb]
\centering
\begin{center}
\vspace{0.0cm}
$\begin{array}{@{\hspace{-0.5in}}c@{\hspace{-0.2in}}c}
\multicolumn{1}{l}{\mbox{}} &
\multicolumn{1}{l}{\mbox{}} \\ [-0.10in]
\epsfxsize=3.8in
\epsffile{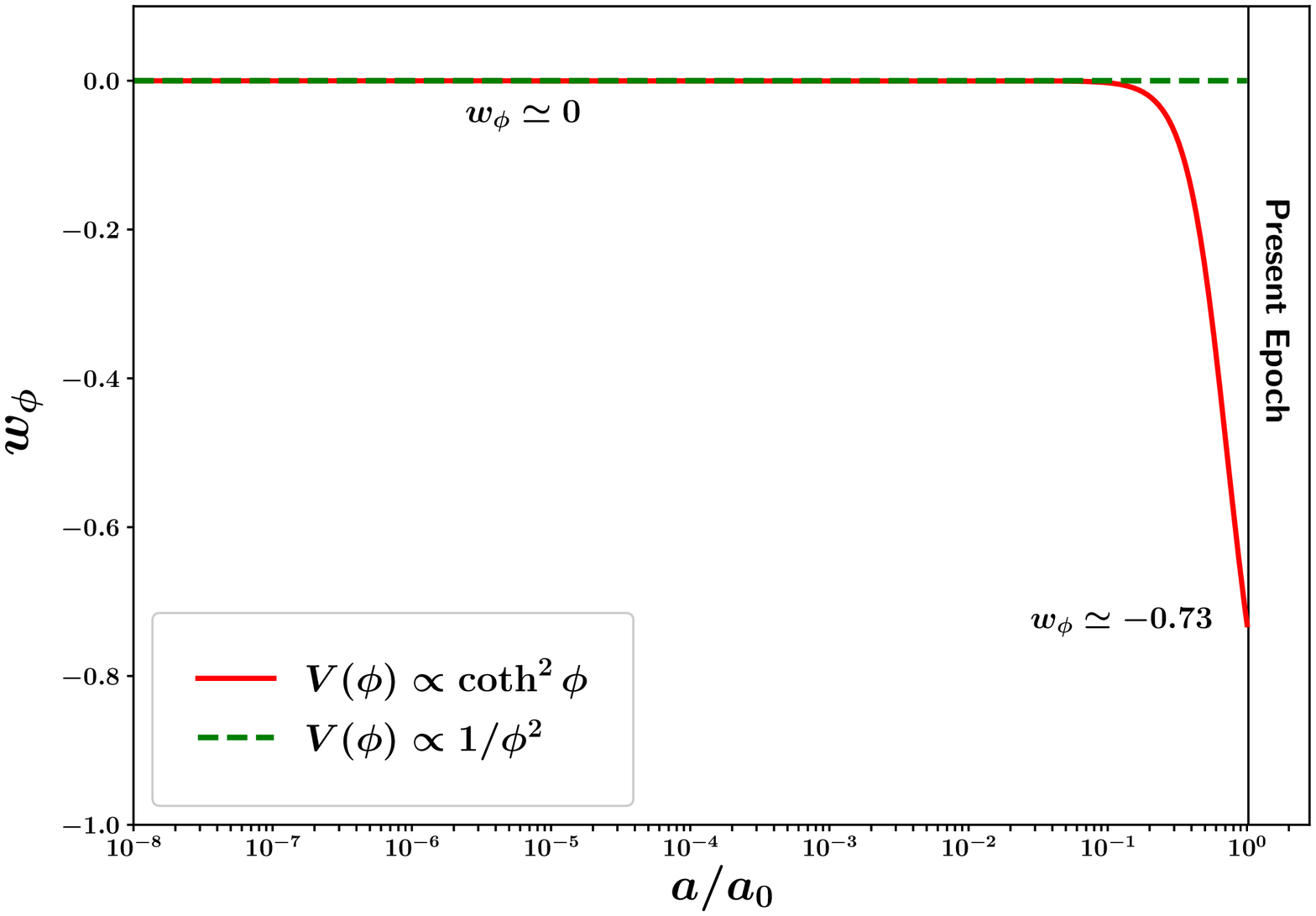} &
\epsfxsize=3.8in
\epsffile{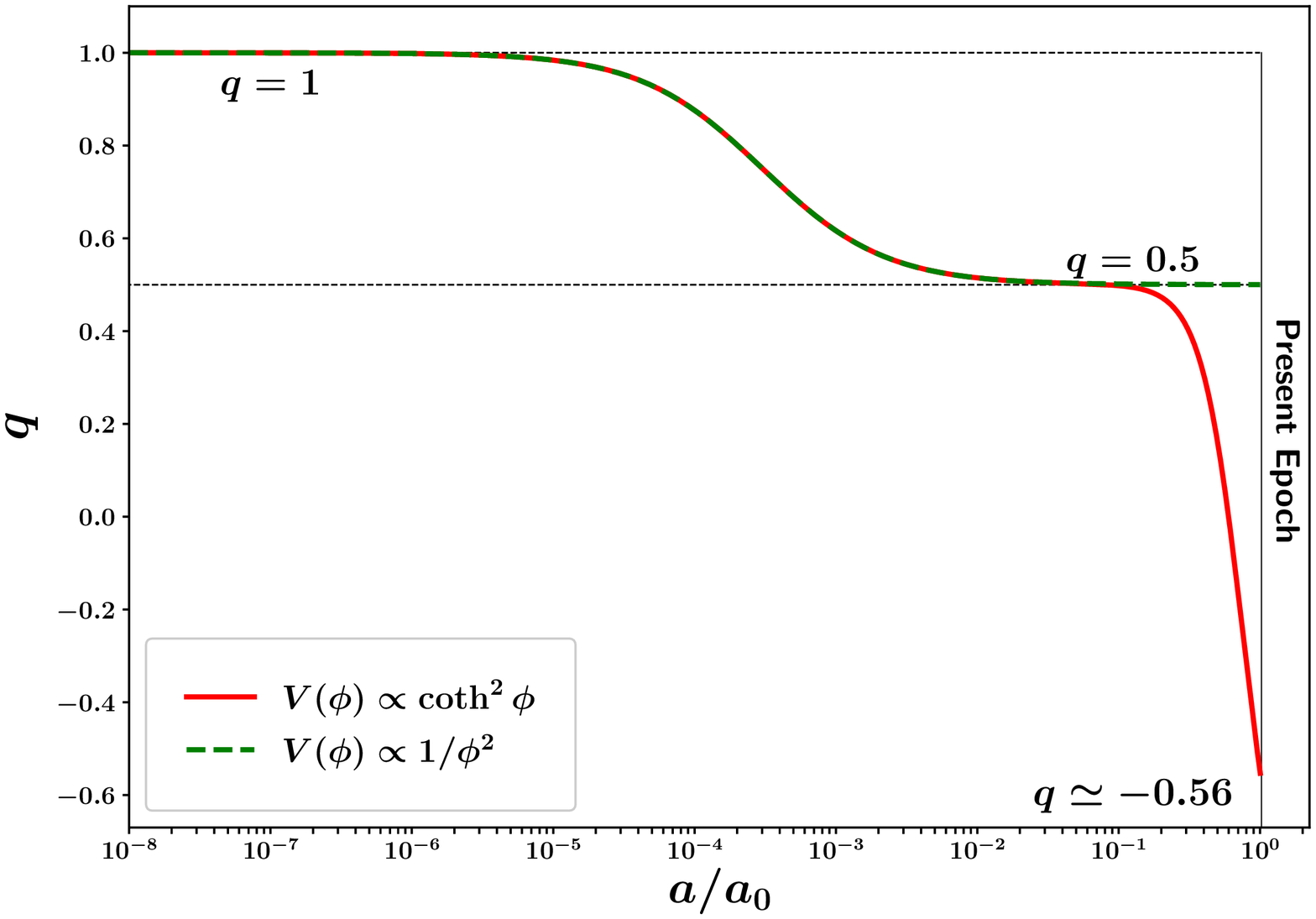} \\
\end{array}$
\end{center}
\vspace{0.0cm}
\caption{\small
The equation of state $w_\phi$ and the deceleration parameter $q$ are shown in the left and right panels
respectively for the potentials
$V(\phi)=V_0\coth^{2}{\left(\frac{\phi}{m_p}\right)}$ (solid red) and 
$V(\phi)=V_0 \left(\frac{m_p}{\phi}\right)^{2}$ (dashed green).
From the {\bf left panel} one finds that $w_{\phi}$ 
in both models remains pegged at $w_{\phi} \simeq 0$ for $z \ggeq 10$. For $z < 10$ the EOS
remains unchanged in IPL but
declines to negative values 
 in the $\coth$ potential. 
This is substantiated by the {\bf right panel} which shows that the deceleration parameter
in both models remains the same until $z \sim 10$. Thereafter $q$
declines to negative values at late times
 for the $\coth$ potential (red curve), 
reflecting the late-time acceleration of the universe.
For the IPL potential on the other hand (dashed green) the universe stays matter dominated at 
late times ($q \simeq 0.5$) and does not accelerate.
}
\label{fig:w_q_coth}
\end{figure}

For non-canonical models the equation of state can be determined from (\ref{eqn:rho}), namely
\beq
w_{\phi}= \frac{p_{_{\phi}}}{\rho_{_{\phi}}} = -1 + \left(\frac{2\alpha}{2\alpha - 1}\right)\left(\frac{\rho_{_X}}{\rho_{_X}+V(\phi)}\right)~,
\label{eq:EOS_nc_1}
\eeq
which simplifies to
\beq
w_{\phi}= -1 + \frac{\rho_{_X}}{\rho_{_X}+V(\phi)}~, ~~~{\rm for}~~\alpha\gg 1~.
\label{eq:EOS_nc_2}
\eeq
For the potential (\ref{eq:coth}) $\rho_{_X} \propto a^{-3}$ whereas $V(\phi)$ approaches a constant value 
at late times.
Consequently one gets 
 \beq
w_{\phi}(z=0)\simeq -1 + \frac{\Omega_{0X}}{\Omega_{0X}+\Omega_{0V}}~,
\label{eq:EOS_nc_0}
\eeq
where $\Omega_{0V} = V_0/\rho_{cr,0}$.
 Assuming $\Omega_{0V}=0.7$, $\Omega_{0b}=0.04$, one finds $\Omega_{0X}=0.26$ so that
 \beq
 w_{\phi}(z=0)\simeq -0.7292.
 \label{eq:EOS_nc_att}
\eeq
Our results for the deceleration parameter $q = -\frac{\ddot a}{aH^2}$ and $w_\phi$
are shown in figure \ref{fig:w_q_coth}
for the IPL model (\ref{eq:IPL}) and the $\coth$ potential (\ref{eq:coth}), both with $p=2$.

The effective equation of state of the kinetic and potential components can be determined from
\ber
w_{_X} &=& - \frac{\dot{\rho_{_X}}}{3H\rho_{_{X}}}-1 ,\label{eq:wX} \\
w_{_V} &=& -\frac{\dot{V}}{3H{V}}-1~ .~~
\label{eq:wV}
\eer
Substituting for  $\rho_{_{X}}$ from \eqref{eqn:rhoNC} one finds 
\beq
w_{_X} = -1-\frac{2\alpha{\ddot\phi}}{3H{\dot\phi}}~.
\eeq
which leads to 
\beq
w_{_X} = c_{_S}^2 =(2\alpha - 1)^{-1}
\eeq
after substitution for $3H{\dot\phi}$ from \eqref{eq:phidot}.

It is instructive to rewrite \eqref{eq:wV} as
\beq
1+w_{_V} = -\frac{\dot{V}}{3H\rho_{_{X}}} \left(\frac{\rho_{_{X}}}{V}\right )~.
\label{eq:wV1}
\eeq
As noted in \eqref{eq:inequality2} the inequality ${\dot V} \ll 3H\rho_{_X}$
 should be satisfied in order
for $\rho_{_{X}}$ to behave like dark matter. Since
the densities in dark matter and dark energy are expected to be comparable at late times,
one finds $\frac{\rho_{_{X}}}{V} \sim O(1)$ at $z \leq 1$. Substituting 
these results 
in \eqref{eq:wV1}
one concludes that the EOS of
 DE is expected to approach $w_{_V} \simeq -1$ at late times in unified models of the dark sector. 
(One should also note that since $\frac{\rho_{_{X}}}{V}$ can be fairly large at early times, 
$w_{_V}$ is not restricted to being close to -1 at $z \gg 1$.)

\begin{figure}[htb]
\centering
\begin{center}
\vspace{0.0cm}
$\begin{array}{@{\hspace{-0.5in}}c@{\hspace{-0.2in}}c}
\multicolumn{1}{l}{\mbox{}} &
\multicolumn{1}{l}{\mbox{}} \\ [-0.10in]
\epsfxsize=3.8in
\epsffile{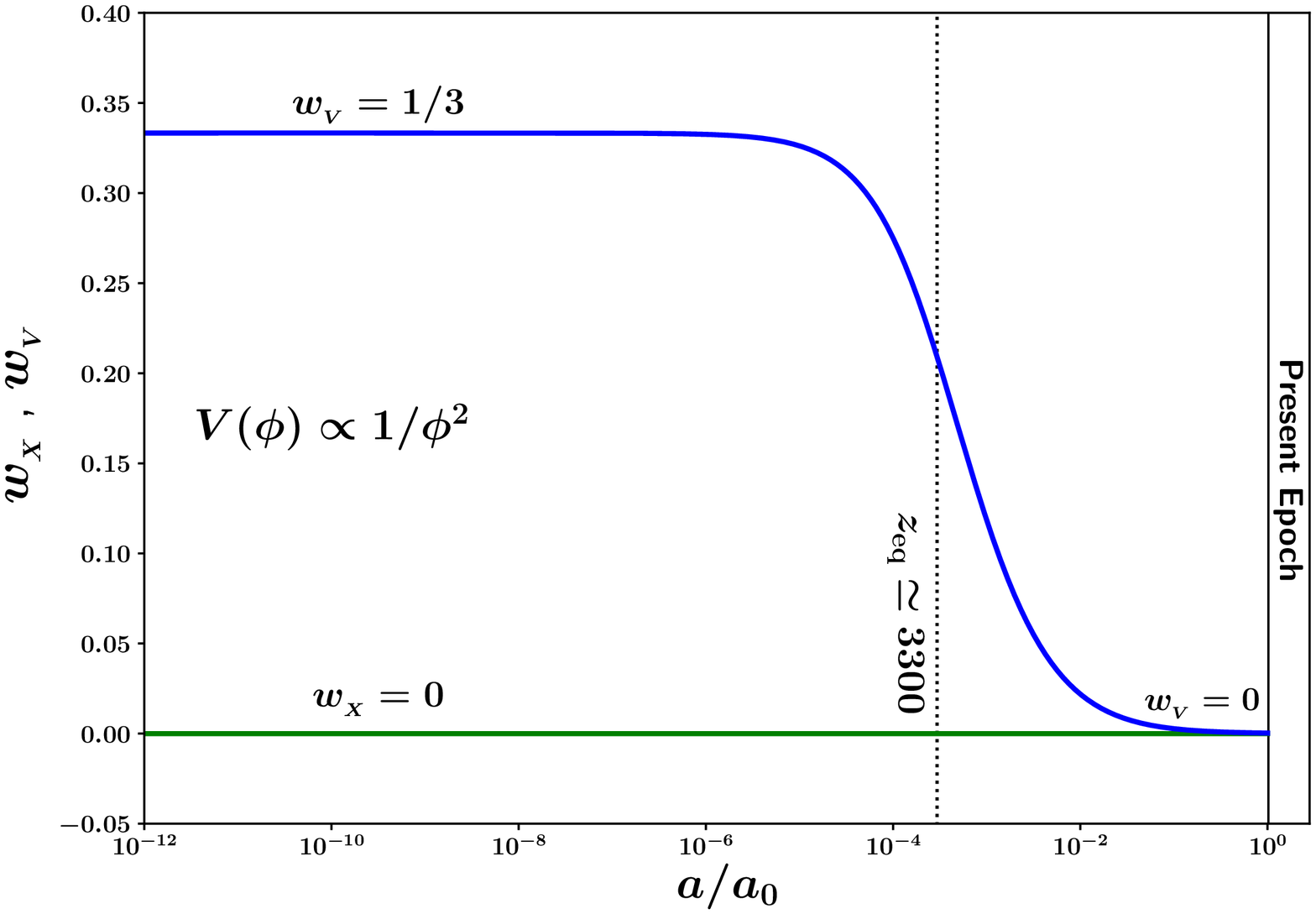} &
\epsfxsize=3.8in
\epsffile{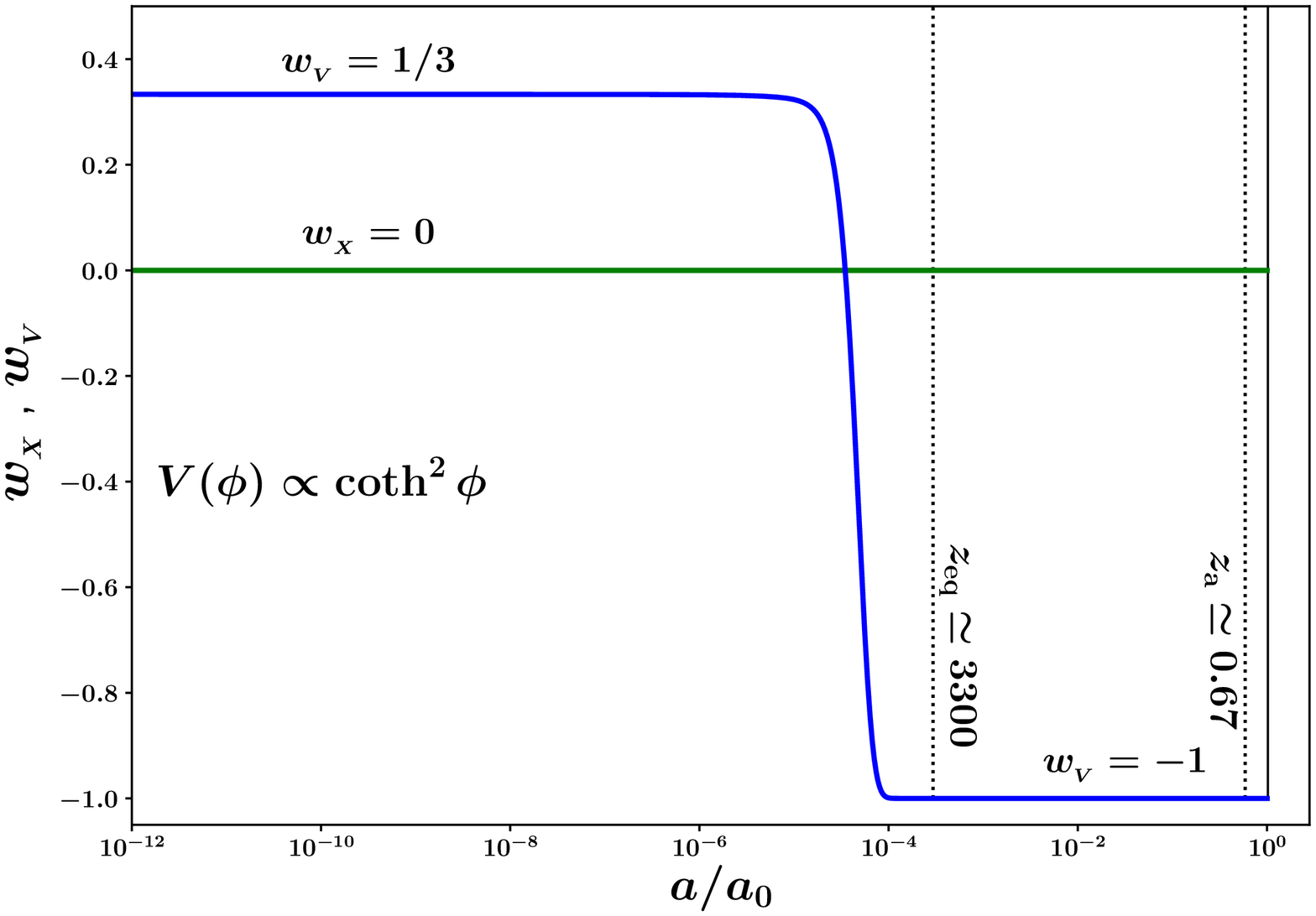} \\
\end{array}$
\end{center}
\vspace{0.0cm}
\caption{\small The evolution of the EOS of the kinetic term  $w_{_{X}}$ (solid green curve) and the potential 
term $w_{_{V}}$ (solid blue curve) is shown for the  IPL potential $V(\phi)=V_0 \left(\frac{m_p}{\phi}\right)^{2}$  
(\textbf{left panel}) and  the coth  potential $V(\phi)=V_0\coth^{2}{\left(\frac{\phi}{m_p}\right)}$ (\textbf{right panel}). 
For the IPL potential, $w_{_{V}}$ scales like the background fluid and  {\em never becomes negative}, 
indicating that the universe enters a prolonged matter dominated epoch at late times. By contrast,
for the coth potential $w_{_{V}}$ behaves like  a scaling solution only at early times 
($w_{_{V}} = w_{_B}$) and drops to  $w_{_{V}}\simeq -1$ at $z \sim 10^4$. 
For both potentials the EOS of the kinetic term stays pegged at $w_{_{X}} = 0$, allowing it to play the role of
dark matter.
}
\label{fig:wXwV_coth_ipl}
\end{figure}

Figure \ref{fig:wXwV_coth_ipl} compares  the behaviour of $w_{_X}$ and $w_{_V}$ in the
two potentials: $\coth^2{\phi}$ and $\phi^{-2}$.
 One notices that $w_{_V}\simeq 1/3$ at early times in both potentials,
which is a reflection of the scaling behaviour $V \propto \rho_{_{\rm B}}$ noted in \eqref{eq:PE_att}.
 At late times $w_{_V}$ in the coth potential drops to negative values causing the universe to accelerate.
 For $V \propto \phi^{-2}$ on the other hand, $w_{_V}$ always tracks the dominant background fluid
which results in $w_{_V} = 0$ at late times. 
(Note that in this case the fluid which dominates at late times is the kinetic term, so that 
$V \propto \rho_{_X}$.)

\subsection{Dark Matter and Dark Energy from a Starobinsky-type potential}
\label{sec:star}

A unified model of dark matter and dark energy can also arise from the potential \cite{linde1,linde2,sss17} 
\beq
V(\phi) = V_0 \left ( 1 - e^{-\lambda\frac{\phi}{m_p}}\right )^{2}~,~~\lambda>0 ,
\label{eq:star1}
\eeq
which reduces to the Starobinsky potential in the Einstein frame \cite{star} for $\lambda=\sqrt{\frac{2}{3}}$.

\begin{figure}[ht]
\centering
\includegraphics[width=0.69\textwidth]{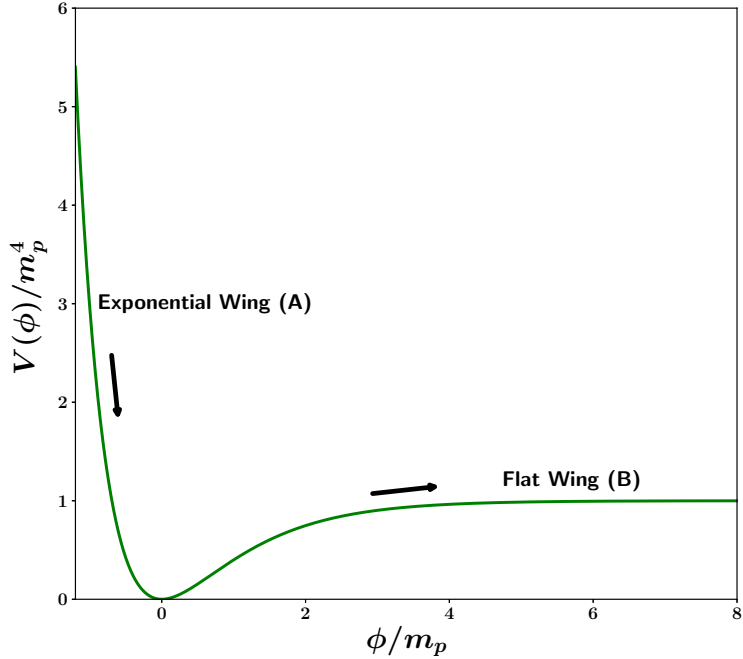}
\caption{\small This figure schematically illustrates the Starobinsky-type potential (\ref{eq:star1})
 with $\lambda=1$.
The main features of this potential are: the exponential wing for $\lambda|\phi| \gg m_p$
($\phi<0$),
the flat wing for $\lambda\phi \gg m_p$, and a  minimum at $\phi=0$, near which
$V \propto \phi^2$. }
\label{fig:staroughpot}
\end{figure}

The potential in \eqref{eq:star1} is characterized by three asymptotic branches (see
figure~\ref{fig:staroughpot}):
\ber
\mbox{Exponential branch:} \quad   V(\phi) &\simeq& V_0\, e^{-2\lambda\phi/m_p}~, \quad
\phi < 0\, , \quad \lambda|\phi| \gg m_p\, ,
\label{eq:starpot1}\\
\mbox{flat branch:} \quad V(\phi) &\simeq& V_0\, , \quad \lambda\phi \gg m_p \, ,
\label{eq:starpot2}\\
\mbox{minimum:} \quad V(\phi) &\simeq& \frac{1}{2}\mu^2\phi^2\, , \quad
\lambda|\phi| \ll m_p \, , \label{eq:starpot3}
\eer
where
\beq
\mu^2 = \frac{2V_0\lambda^2}{m_{p}^2}.
\eeq

Before discussing the unification of dark matter and dark energy in a Starobinsky-type potential
we briefly explore the dynamics of the scalar field as it rolls down the exponential branch (\ref{eq:starpot1}).

\subsubsection{Motion along the exponential branch}
\label{sec:expo}

\begin{figure}[htb]
\centering
\includegraphics[width=0.8\textwidth]{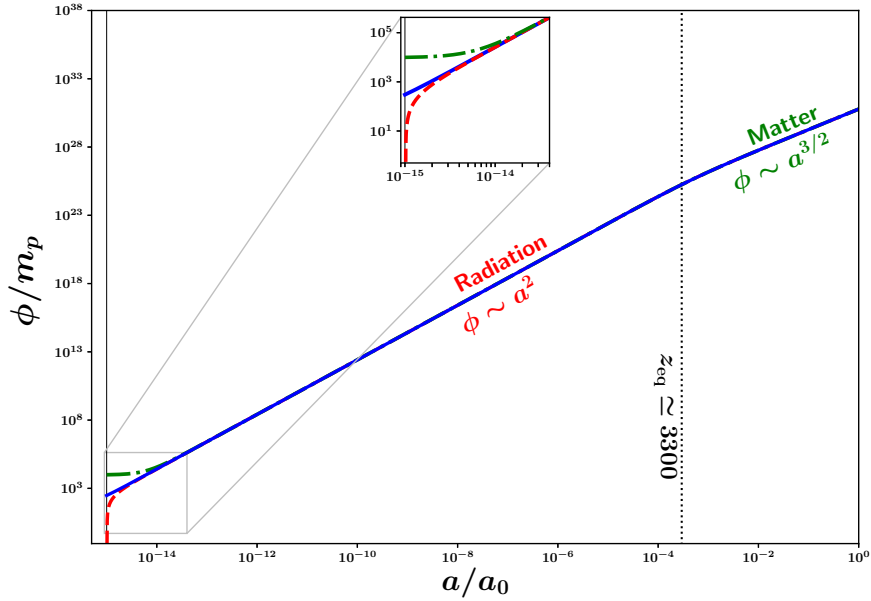}
\caption{\small The evolution of $\phi$  is shown for the exponential potential  
$V(\phi)=V_0~ e^{-\frac{2\phi}{m_p}}$. Commencing at $z \sim 10^{15}$, 
different initial conditions (represented by dashed green and red curves) rapidly
  converge onto the attractor solution (\ref{eq:phi_star}) (solid blue curve) which 
corresponds to
  $\phi \propto a^2$ during radiation domination and $\phi \propto a^{3/2}$ during matter domination.
The rapid growth in $\phi$ is accompanied by a steep decline in $V(\phi)$.}
\label{fig:phi_att_exp}
\end{figure}

The exponential branch has been extensively studied in the canonical case ($\alpha=1$)
for which the late time attractor is $w_\phi = w_{_{\rm B}}$ if
$\lambda^2 > 3(1+w_{_{\rm B}})$. 
The situation radically changes for non-canonical scalars ($\alpha \neq 1$). 
As shown in \cite{li-scherrer}, if the background density scales as $\rho_{_{\rm B}} \propto a^{-m}$, then for 
$(2\alpha - 1) m > 6$
the late time attractor is
\beq
{\dot\phi}^{2\alpha - 1} \propto t^{-6/m}
~~~ \Rightarrow ~~ {\dot\phi} \propto a^{-\frac{3}{2\alpha - 1}}~.
\label{eq:attractor}
\eeq
Since $\rho_{_X} \propto {\dot\phi}^{2\alpha}$ one finds
\beq
\rho_{_X} \propto a^{-\frac{6\alpha}{2\alpha - 1}}
\eeq
which reduces to
\beq
\rho_{_X} \propto a^{-3} ~~~ {\rm for}~~ \alpha \gg 1.
\eeq
We therefore find that, as in the IPL case, for large values of the non-canonical parameter $\alpha$
the density of the kinetic term scales just like pressureless (dark) matter.
From \eqref{eq:attractor} one also finds that ${\dot\phi} \sim$ {\em constant} when 
$\alpha \gg 1$. From this it 
is easy to show that for $\alpha \gg 1$ the amplitude of the scalar field grows as 
\beq
\frac{\phi}{m_p} \propto a^{m/2}~,
\label{eq:phi_star}
\eeq
so that $\phi \propto a^2$ during the radiative regime and $\phi \propto a^{3/2}$ during matter domination.
This behaviour is illustrated in figure \ref{fig:phi_att_exp}.
Substituting \eqref{eq:phi_star} in  (\ref{eq:starpot1}) one finds 
$V \propto \exp{[-2\lambda\phi]} \sim \exp{[-2\lambda a^{m/2}]}$, which implies
an exponentially rapid decline in the value of the potential as the universe expands and $a(t)$ increases. 
One therefore concludes
that like the IPL potential,
 an exponential potential too can never dominate the energy density of the universe and 
source cosmic acceleration.

\subsubsection{Accelerating Cosmology from a Starobinsky-type potential}
\label{sec:star_acc}

\begin{figure}[htb]
\centering
\begin{center}
\vspace{0.0cm}
$\begin{array}{@{\hspace{-0.5in}}c@{\hspace{-0.2in}}c}
\multicolumn{1}{l}{\mbox{}} &
\multicolumn{1}{l}{\mbox{}} \\ [-0.10in]
\epsfxsize=3.8in
\epsffile{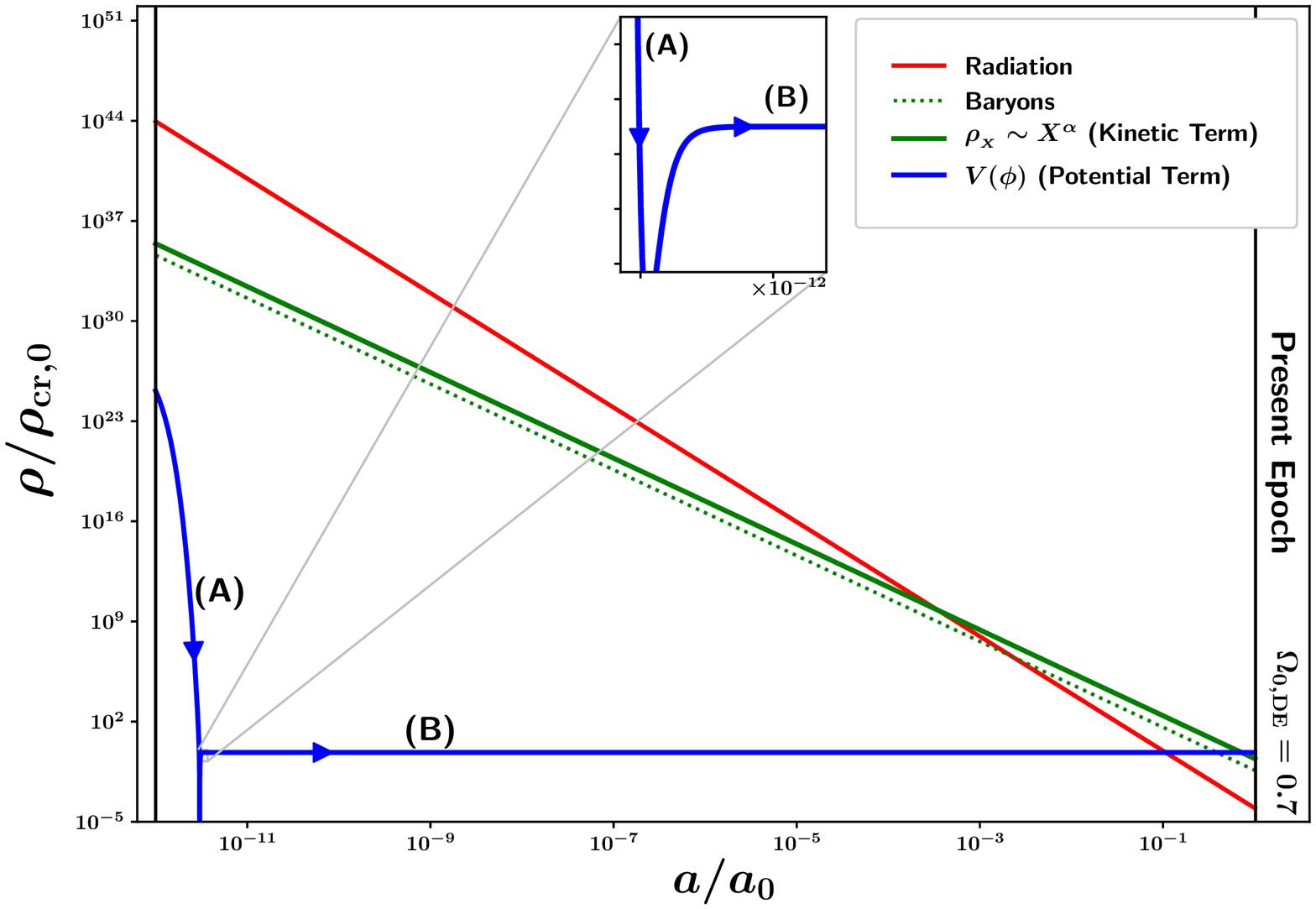} &
\epsfxsize=3.8in
\epsffile{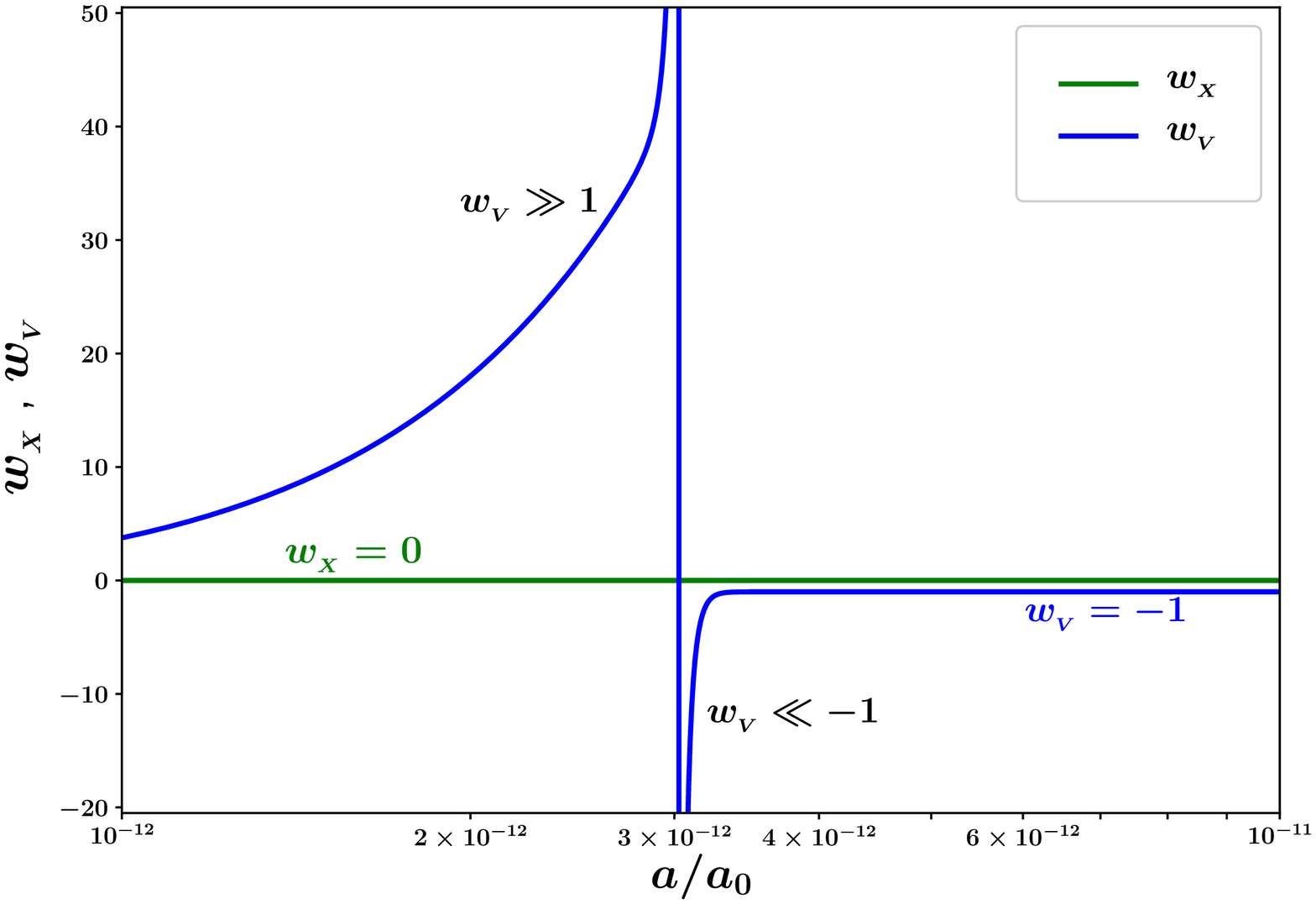} \\
\end{array}$
\end{center}
\vspace{0.0cm}
\caption{\small {\bf Left panel}: The density of radiation (red curve), baryons (dotted green curve) 
and the non-canonical scalar field 
are shown for the Starobinsky-type potential (\ref{eq:star1}) with  $\lambda=1$.
The solid green curve shows the evolution of the kinetic term $\rho_{_X} \propto X^\alpha$ 
which behaves like dark matter throughout the history of the universe. The solid  blue curve shows
 the  evolution of the potential $V(\phi)$. Note that $V(\phi)$ rapidly decreases 
as $\phi$ rolls down the exponentially steep left wing (A) of the potential. This results in 
$V(\phi)$ falling to zero (at $\phi\simeq 0$)  and then increasing to $V_0$, as $\phi$ 
commences to roll along the flat right wing (B); see fig.\ref{fig:staroughpot}. 
Since $\rho_{_X} \propto a^{-3}$, the late time expansion in this model resembles $\Lambda$CDM.
{\bf Right panel}: The equation of state of the kinetic term  ($w_{_{X}}$) and the potential term ($w_{_{V}}$)   
are shown for the transition region when the scalar field rolls from the steep left wing (A)
onto the flat right wing (B) of the potential (\ref{eq:star1}). 
 The change in slope of the
potential at $\phi \simeq 0$ leads to
a pole-like singularity in $w_{_{V}}$.
(Note that $w_{_{V}}$ as defined in \eqref{eq:wV} is an effective quantity hence its value can exceed unity.)
By contrast the EOS of state of the kinetic term remains pegged at $w_{_{X}}=0$.
}
\label{fig:rho_staro}
\end{figure}

In the context of the Starobinsky-type potential in \eqref{eq:star1},
the rapid growth of $\phi$ in (\ref{eq:phi_star}) enables the scalar field to pass from the steep left wing to the 
flat right wing of $V(\phi)$.
In other words the scalar field rolls from A to B in figure \ref{fig:staroughpot}.
Since $V \simeq V_0$ on the flat right wing, 
cosmological expansion in this model mimicks $\Lambda$CDM at late times.
This is illustrated in figure \ref{fig:rho_staro}.
(Note that $\rho_{_X} \propto a^{-3}$ on both wings of the potential.)

Figure \ref{fig:rho_staro} shows the behaviour of $w_{_X}$ and $w_{_V}$ 
as $\phi$ moves under the influence of 
the potential 
\eqref{eq:star1}. A key feature to be noted is that
 $w_{_V}$ encounters a {\em pole} as $\phi$ rolls from the steep
left wing to the flat right wing of $V(\phi)$ (\ie from A to B in figure \ref{fig:staroughpot}).

\subsection{Cascading Dark Energy}
\label{sec:cascade}

\begin{figure}[htb]
\centering 
\begin{center}
\vspace{0.0cm}
$\begin{array}{@{\hspace{-0.5in}}c@{\hspace{-0.2in}}c}
\multicolumn{1}{l}{\mbox{}} &
\multicolumn{1}{l}{\mbox{}} \\ [-0.10in]
\epsfxsize=3.8in
\epsffile{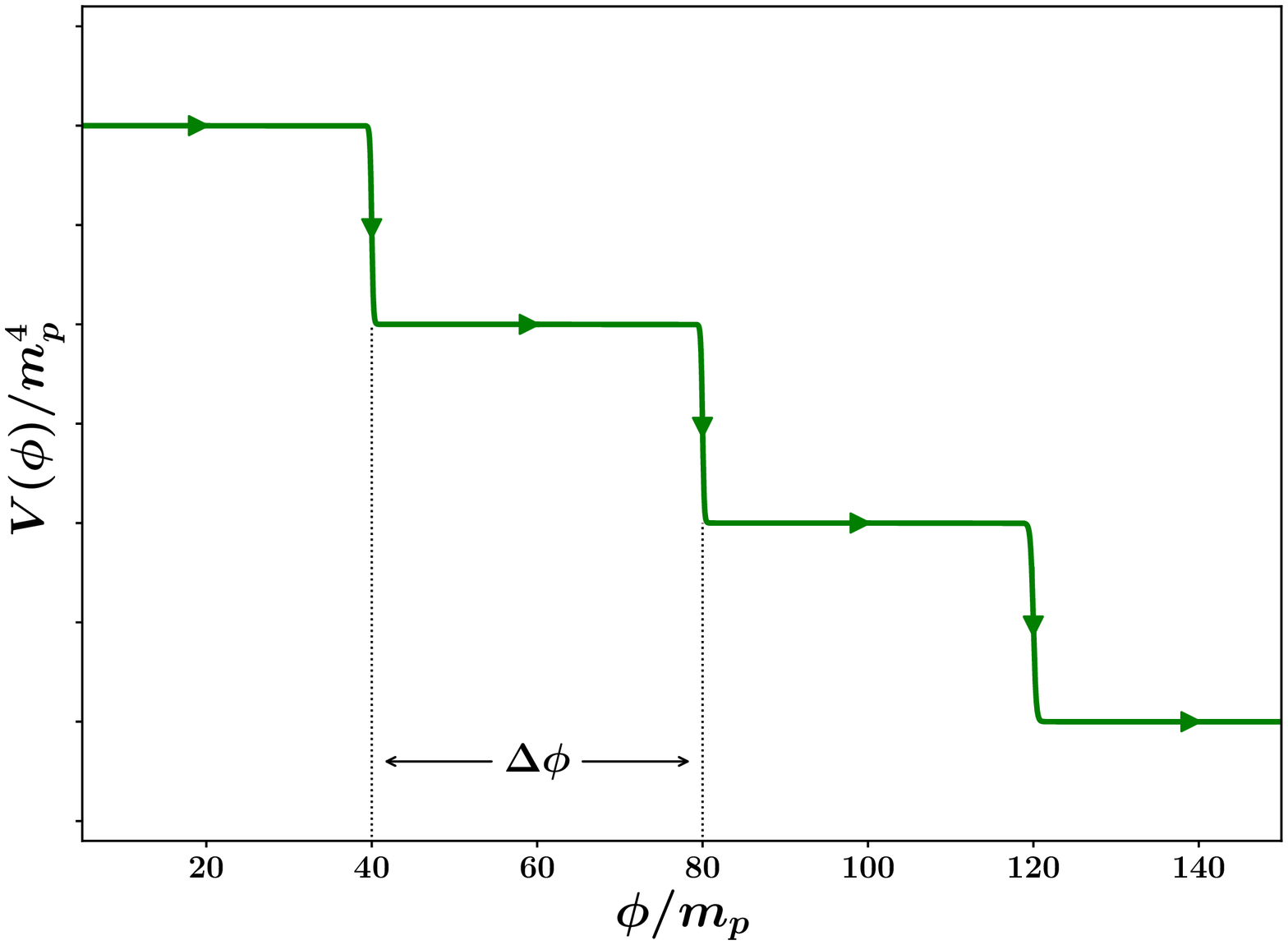} &
\epsfxsize=3.8in
\epsffile{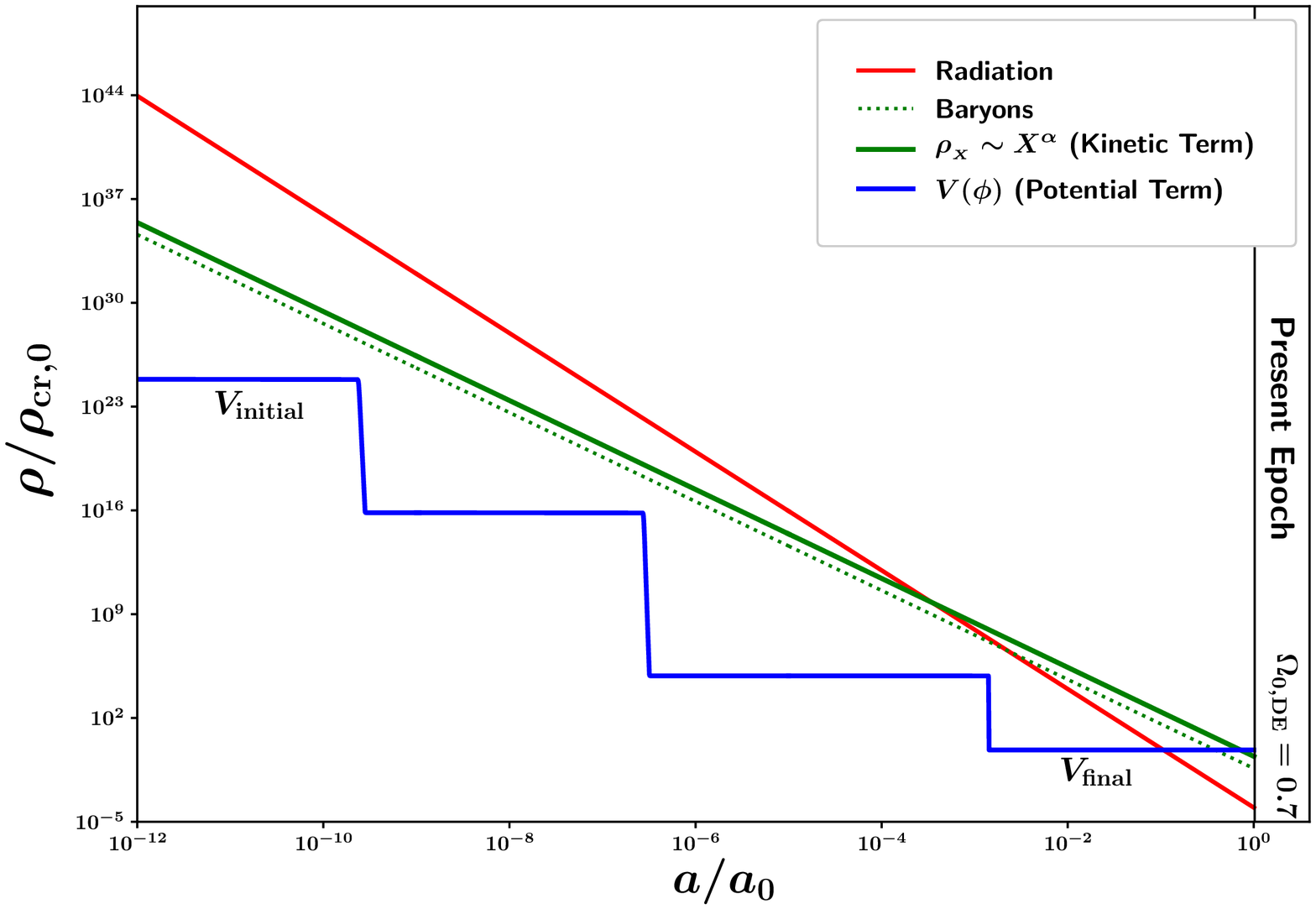} \\
\end{array}$
\end{center}
\vspace{0.0cm} 
\caption{\small
A schematic view of the cascading potential is shown in the {\bf left panel}.
The {\bf right panel} shows the evolution in the density of the kinetic term 
$\rho_{_X} \propto a^{-3}$ (solid green line) and the cascading potential $V(\phi)$ (blue).
Also shown are the densities in radiation (red) and baryons (dotted green).
(Note that $V_{\rm initial} = 10^{25}\times V_{\rm final}$.)
}
\label{fig:cascade}
\end{figure}

A question occasionally directed towards dark energy is whether cosmic acceleration will continue forever
(as in $\Lambda$CDM) or whether, like the earlier transient epochs
(inflation, radiative/matter dominated) dark energy will also will be
a fleeting phenomenon.
In this section we investigate a transient model of DE in which the potential $V(\phi)$ 
is piece-wise flat and resembles
a staircase; see figure \ref{fig:cascade}.
Such a potential might mimick a model in which an initially large vacuum energy cascades to lower values through
a series of waterfalls -- discrete steps \cite{watson06}.

Locally the i-th step of this potential may be described by \cite{sahni_sen,tower09}
\beq
V(\phi) = A + B\tanh{\beta \phi}
\label{eq:step}
\eeq
where $A+B = V_{\rm i}$
and $A-B = V_{\rm i+1}$.
If $V_{\rm i+1} \simeq 10^{-47} {\rm GeV}^4$ then this potential could account for cosmic
acceleration. (One might imagine yet another step at which
$V_{\rm i+2} < 0$. In this case the universe would stop expanding and begin to 
contract at some point in the future.)
Motion along the staircase potential leads to a cascading model of dark energy.
Remarkably, the inequality in \eqref{eq:inequality2}
 holds even as $\phi$ cascades from higher to lower values of $V$.
This ensures that the kinetic term scales as $\rho_{_X} \propto a^{-3}$ and behaves
like dark matter while $V(\phi)$ behaves like dark energy, as shown in fig. \ref{fig:cascade}. 

Note that the cascading DE model runs into trouble in the canonical context since 
the kinetic energy of a canonical scalar
field moving along a flat potential declines as $\frac{1}{2}{\dot\phi}^2 \propto a^{-6}$.
This puts the brakes on $\phi(t)$ which soon approaches its asymptotic value $\phi_*$,
resulting in inflation sourced by $V(\phi_*)$.
By contrast ${\dot\phi} \sim$ {\em constant} in non-canonical models with $\alpha \gg 1$,
see \eqref{eq:attractor}.
This allows $\phi(t)$ to cross each successive step on the DE staircase in a finite amount of time,
$\Delta t \simeq \frac{\Delta\phi}{\dot\phi}$, and drop to a lower value of $V(\phi)$; see left panel
of fig. \ref{fig:cascade}.

\section{Discussion}
In this paper we have demonstrated that a scalar field with a non-canonical kinetic term can play the
dual role of dark matter and dark energy. 
The key criterion which must be satisfied by unified models of the dark sector
is \eqref{eq:inequality2}. This inequality ensures that the third term in the equation of motion 
\eqref{eqn: EOM-model}
is small and can be neglected, resulting in $\rho_{_X} \propto a^{-3}$ and $w_{_X} \simeq c_{_S} \simeq 0$.
In other words if \eqref{eq:inequality2} is satisfied the kinetic term behaves like dark matter with vanishing
pressure and sound speed. Of equal importance is the fact that if
 \eqref{eq:inequality2} holds then eqn. \eqref{eq:wV1} implies 
$w_{_V} \simeq -1$ at late times. This ensures that the potential $V(\phi)$ can dominate over $\rho_{_X}$ 
and source cosmic acceleration at late times.

The following unified models of the dark sector have been discussed in this paper:

(i) Models with exactly flat potentials $V' = 0, ~V=V_0$.
As shown in \cite{sahni_sen} the entire expansion history of this model resembles $\Lambda$CDM.
(ii) Successful unification of the dark sector can also arise from
potentials which are steep at early
times and flatten out at late times. 
Both the $\coth$ potential \eqref{eq:coth} and the Starobinsky-type potential \eqref{eq:star1}
 provide us with examples of this category.
(iii) The step-like potential \eqref{eq:step} also leads to unification
\footnote{Note that the width of each step is restricted by the fact that the universe does not
appear to accelerate prior to $z \sim 1$ (with the exclusion of an early inflationary epoch).}.
In this case the motion of $\phi$ resembles a series of waterfalls as $V(\phi)$ cascades to lower
and lower values.
It is interesting to note that
for all of the above potentials \footnote{We note in passing that the
asymptotically flat potential \cite{linde1} $V(\phi) = V_0 \tanh^2{\phi}$
also leads to a unified scenario of dark matter and dark energy although we do not discuss it in this paper.}
 the kinetic term scales as $\rho_{_X} \propto a^{-3}$ 
throughout the expansion history of the universe, 
even as the shape of the potential continuously changes.
This property allows the kinetic term to
play the role of dark matter while
the potential term $V(\phi)$ plays the role of dark energy and leads to cosmic acceleration at late times.

As shown in \cite{sahni_sen} the small (but non-vanishing) speed of sound
in non-canonical models suppresses gravitational clustering
on small scales. 
Non-canonical models with $c_s \ll 1$ can therefore
have a macroscopic Jeans length which might help in
resolving the cusp--core and substructure problems which afflict the standard cold dark
matter scenario.
In this context the dark matter content of our model shares similarities
with warm dark matter \cite{neutrino,neutrino1} and fuzzy cold dark matter \cite{hu00,sahni_wang,Witten,sss17}
both of which are known to possess a large Jeans scale.

Finally it is interesting to note that all of the potentials discussed in this paper
belong to the $\alpha$-attractor family of potentials \cite{linde1,linde2}
and lead to interesting models of inflation, dark matter and dark energy \cite{sss17,bms17}
in the canonical case.

\section*{Acknowledgements}
S.S.M thanks the Council of Scientific and Industrial Research (CSIR), India, for
financial support as senior research fellow. S.S.M also acknowledges the hospitality of the
  Yukawa Institute for Theoretical Physics (YITP) at Kyoto University, where a substantial part of this research work was conducted during the workshop "Gravity and Cosmology 2018".

\end{document}